\documentclass[28pt,a4paper]{article}
\usepackage[top=2.5cm, bottom=2.5cm, left=2.5cm, right=2.5cm]{geometry}
\usepackage{mathrsfs}
\usepackage{amsfonts}
\usepackage{amssymb}
\usepackage{graphicx}
\usepackage{amsmath}
\usepackage{extarrows}
\usepackage{arydshln}
\newcommand{\ud}{\mathrm{d}}
\newcommand{\ue}{\mathrm{e}}
\newcommand{\ui}{\mathrm{i}}
\renewcommand{\[}{\begin{equation}}
\renewcommand{\]}{\end{equation}}

\renewcommand{\Im}{\mathrm{Im}}

\newcommand{\Tr}{\mathrm{Tr}}
\makeatletter
\newcommand\figcaption{\def\@captype{figure}\caption}
\renewcommand{\b}[1]{\mathbf{#1}}

\begin{document}
\vspace{2cm}

\title{\Large \bf Spin Generation Via Bulk Spin Current in Three Dimensional Topological Insulators}
\author{Xingyue Peng$^\textbf{1}$,\ Yiming Yang$^\textbf{1}$,\ Rajiv R. P. Singh$^\textbf{1}$,\ Sergey Y. Savrasov$^\textbf{1}$ and Dong Yu$^\textbf{1,2}$}

\makeatletter
\footnotetext[1]{Department of Physics, University of California, Davis, One Shields Avenue,
Davis, California 95616 USA}
\footnotetext[2]{Email: yu@physics.ucdavis.edu}
\maketitle

\begin{abstract}
To date, spin generation in three-dimensional topological insulators is primarily modeled as a single-surface phenomenon, attributed to the momentum-spin locking on each individual surface. In this article we propose a mechanism of spin generation where the role of the insulating yet topologically non-trivial bulk becomes explicit: an external electric field creates a transverse pure spin current through the bulk of a three-dimensional topological insulator, which transports spins between the top and bottom surfaces. Under sufficiently high surface disorder, the spin relaxation time can be extended via the Dyakonov-Perel mechanism. Consequently both the spin generation efficiency and surface conductivity are largely enhanced. Numerical simulation confirms that this spin generation mechanism originates from the unique topological connection of the top and bottom surfaces and is absent in other two dimensional systems such as graphene, even though they possess a similar Dirac cone-type dispersion.
\end{abstract}

\newpage

Topological insulators (TIs) have attracted world-wide attention because of their intriguing fundamental physics and exciting application opportunities in spintronics \cite{colloquium}. Three dimensional (3D) TIs \cite{3dti,3dti_exp} are of particular technological importance since the unique spin generation can be realized in single crystals rather than in complex heterogeneous structures \cite{HgTe_exp}. TIs are considered as efficient spin generators \cite{Mc_review}, yet the spin generation is generally regarded as a pure surface phenomenon. Namely, the electronic momentum and spin are locked at the TI surface, and a net charge current leads to a net spin polarization at the surface, whose magnitude is directly proportional to the charge current \cite{sp_lockin_exp}. In this view, all physics occur independently at the top and bottom surfaces of a TI and the role of the bulk is passive which simply separates the top and bottom surfaces. The surface conductivity is understood through density of states (DOS) and scattering rate, just like in other 2D systems such as graphene and two-dimensional electron gas (2DEG). The conductivity behavior governs the spin generation on the surface of a 3D TI, and spin accumulation is merely a side product of conductivity. While this interpretation of spin generation in TIs is most mathematically straightforward, it is far from satisfactory in the sense that the most amazing feature of a TI -- surface-bulk correspondence does not explicitly enter this physical picture.

\begin{figure}[!hbp]
\begin{center}
\includegraphics[width=\textwidth]{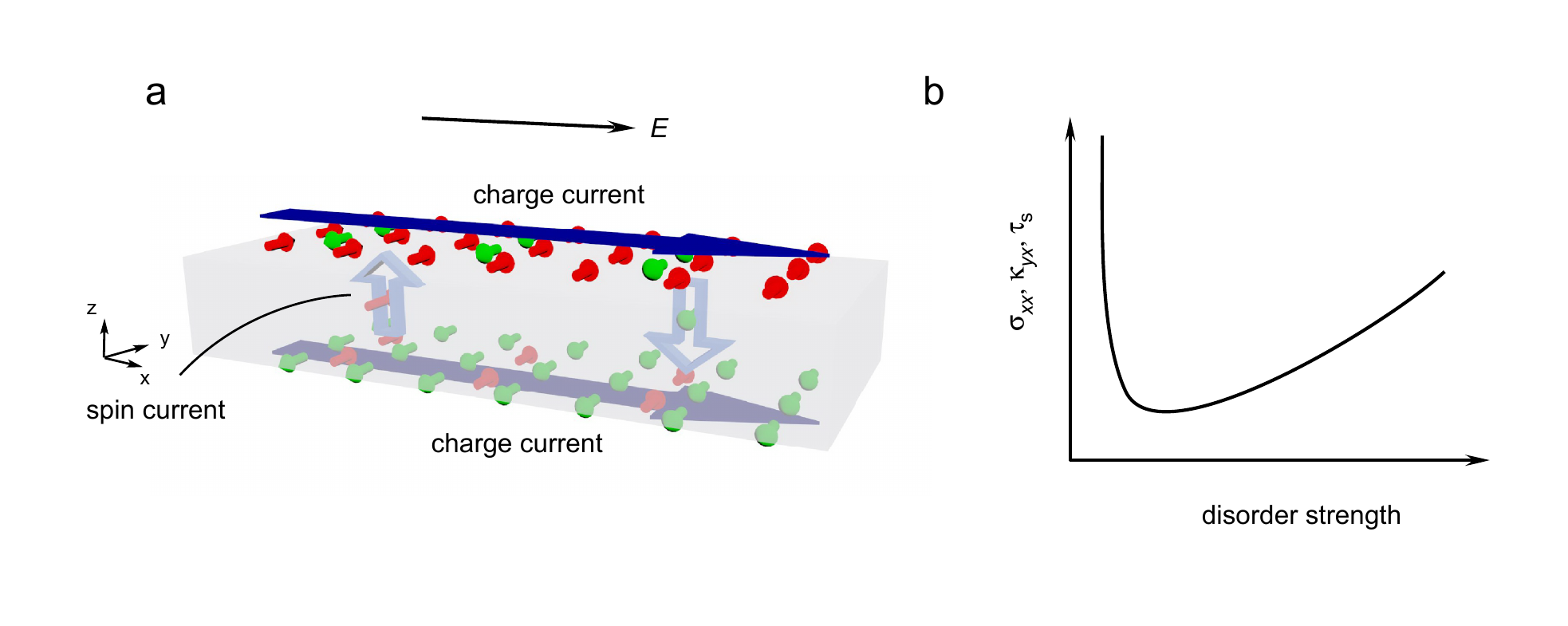}
\caption{Proposed spin dynamics in a 3D TI. {\bf a,} An electric field induces a transverse pure spin current in the bulk. Consequently, opposite spins accumulate on the top and bottom surfaces, leading to a charge current according to the chiral momentum-spin texture. The small cylindrical arrows denote spins. The hollow vertical arrows indicate spin current. The long horizontal blue arrows indicate charge current. {\bf b,} The anomalous behavior of transport coefficients proposed in this article. At a sufficiently high disorder level, conductivity $\sigma_{xx}$, electro-spin susceptibility $\kappa_{yx}$ and spin relaxation time $\tau_{\mathrm{s}}$ should all have positive dependence on the disorder, in contrast with the well known negative dependence in the low disorder limit. }\label{fig:1}
\end{center}
\end{figure}

On the other hand, there is an alternative viewpoint of spin generation. The external electric field induces a transverse pure spin current through the bulk, which acts as a bridge for transporting spins between top and bottom surfaces. Opposite spins are thus accumulated on the two surfaces which lead to charge current in the same direction of the electric field due to the opposite chirality of the momentum-spin textures on the top and bottom surfaces (Figure \ref{fig:1}a). An empirical formula for the bulk spin current can be written down as
\[
j^{\mathrm{s}}_{ij}=\sum_k \sigma^{\mathrm{s}}_{ijk}E_k
\]
where $j^{\mathrm{s}}$ is the spin current density, $E$ is the electric field, and $\sigma^{\mathrm{s}}$ is the spin Hall conductivity tensor \cite{QSHE_review}. A system which is electrically insulating but can carry a pure spin current is termed a spin Hall insulator \cite{SHI}. The bulk of a 3D TI has been demonstrated to be a spin Hall insulator due to its $Z_2$ topological order \cite{QSHE_TI}.

Analogous to Hall effect, the transverse spin Hall current leads to surface spin accumulation in a slab geometry.
Yet unlike electric charge, spin is usually a nonconserved quantity in a spin Hall insulator.  The ultimate spin accumulation induced on the surface closely depends on the spin relaxation mechanism.  In the low disorder limit $\mu\tau/\hbar\gg1$ with $\mu$ being the Fermi level, $\tau$ being the momentum relaxation time, it has been demonstrated that the spin relaxation time $\tau_{\mathrm{s}}$ on the surface of a 3D TI is identical to the momentum relaxation time $\tau$ due to the momentum-spin locking, and the traditional Dyakonov-Perel spin relaxation is absent \cite{srelax}. Charge-spin dynamics in the high disorder limit $\mu\tau/\hbar\lesssim1$, however, has rarely been discussed in the literature so far. 

The exact behavior of these spin transport coefficients under high disorder is of crucial importance to the application of 3D TI-based spintronic devices, because unlike the bulk, surface is extremely vulnerable to various kinds of defects, especially when placed in ambient environment. Even for a material which is generally considered ``inert'', the top most layer of atoms could still suffer from high concentration of impurities \cite{inert_surface,surf_oxi}. A clear physical model of charge and spin transport in this case is highly desired for the design of novel 3D TI-based spintronic devices.

In this article, we demonstrate that sufficiently high nonmagnetic disorder can suppress spin relaxation and result in an increase of the spin relaxation time $\tau_{\mathrm{s}}$ in a manner similar to the traditional Dyakonov-Perel mechanism. Consequently, both electro-spin susceptibility $\kappa_{yx}$ (surface spin density $s_y$ divided by the electric field $E_x$) and the electric conductivity $\sigma_{xx}$ should increase with the increase of disorder, as illustrated in Figure \ref{fig:1}b.

\begin{figure}[!hbp]
\begin{center}
\includegraphics[width=\textwidth]{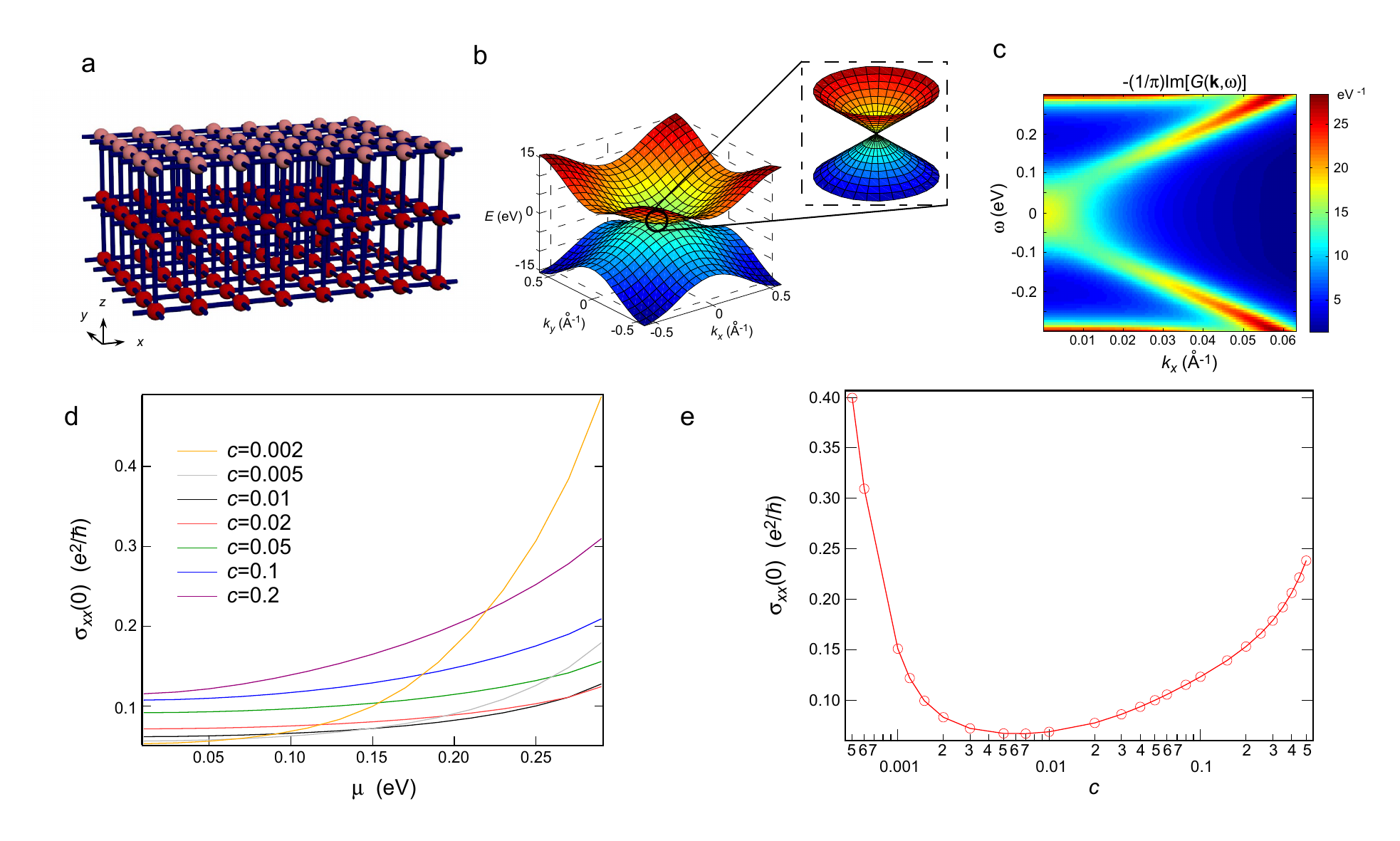}
\caption{Model configurations and conductivity simulation. {\bf a,} The slab of tetragonal lattice is infinite in the $xy$ directions and has a total number of $N=10$ layers in the $z$ direction (c-axis of the lattice). Only the top three layers are shown here. There are four states on each site and only nearest neighbor hopping is considered. The top and bottom layers are subject to vacancies, which are colored pink in this figure. Each atom on these layers has a probability of $c$ to be occupied by a vacancy and $(1-c)$ to be intact. The on-site energy of a vacancy is brought to infinity to forbid electrons from entering this site. {\bf b,} The energy dispersion of the surface branch of a clean system. A Dirac cone exists around the $\Gamma$ point. Parameters used for simulation: $A=1\ \mathrm{eV}$, $A_z=0.5\ \mathrm{eV}$, $B=2\ \mathrm{eV}$, $B_z=0.4\ \mathrm{eV}$, $\Delta=0.3\ \mathrm{eV}$, $a=5\ \mathrm{\AA}$. {\bf c,} The spectral function $-(1/\pi)\Im G(\b{k},\omega)$ obtained via CPA plotted along the $\Gamma-\mathrm{X}$ line at impurity concentration $c=0.001$. At higher concentrations,  the $\b{k}$-dispersion first fades away and then slowly recovers, as discussed in \cite{strong_disorder}. {\bf d,} The DC conductivity of a single surface ($\sigma_{xx}(0)$) plotted against the Fermi level position ($\mu$). The impurity concentration $c$ varies from $0.002$ to $0.2$. {\bf e,} Conductivity ($\sigma_{xx}(0)$) plotted against the impurity concentration ($c$). The Fermi level position was fixed at $\mu=0.13\ \mathrm{eV}$.  }\label{fig:2}
\end{center}
\end{figure}

To begin with, we consider a realistic 4-band tight-binding model \cite{strong_disorder} built on a slab of a tetragonal lattice, as shown in Figure \ref{fig:2}a.
The slab is infinite in $xy$ directions and has a total number of $N=10$ layers in the $z$ direction (c-axis).  With 4 states on each site, the bulk Hamiltonian in the 3D $\b{k}$-space is
\[\label{eq:1}
H_0(\b{k})=\left(A\sum_{i=x,y} \sin k_i a\alpha_i +A_z \sin k_z a_z \alpha_z\right) +\left[\Delta-4\left(B\sum_{i=x,y} \sin^2\frac{k_ia}{2}+B_z \sin^2\frac{k_za_z}{2}\right)\right]\beta
\]
where $\alpha_i(i=x,y,z),\beta$ are the Dirac matrices, $a$ and $a_z$ are the lattice constants in the $xy$ and $z$ directions, $\Delta$ is the mass term and $A,A_z,B,B_z$ are nearest neighbor hopping amplitudes. In the slab configuration, inverse Fourier transform is performed in $z$-direction to comply with the finite thickness.

We use a typical 3D TI $\mathrm{Bi}_2\mathrm{Se}_3$ as our prototype and adopt parameters as obtained in \cite{3dti} to best fit the band structure of $\mathrm{Bi}_2\mathrm{Se}_3$.  The resultant band structure of the surface is shown in Figure \ref{fig:2}b which clearly has a Dirac cone near the $\Gamma$ point. Due to the $z$-inversion symmetry of the slab, all bands are doubly degenerate.

To account for surface disorder, atoms in the top and bottom layers of the slab are subject to a typical kind of impurity -- vacancies. Each site at the surface has a probability of $c$ to be occupied by a vacancy where the on-site energy is brought to infinity so as to forbid electrons from this site. As $c$ may not be small, the first Born approximation does not apply. Here we adopted the coherent potential approximation (CPA) method for binary alloys \cite{CPA,CPA_review} in computing the Green's function $G(\b{k},\omega)$ and self-energy $\Sigma(\omega)$. A typical spectral function $-(1/\pi)\Im G(\b{k},\omega)$ obtained by CPA at impurity concentration $c=0.001$ is plotted in Figure \ref{fig:2}c. The evolution of the spectral function with increasing impurity concentration is consistent with results obtained in \cite{strong_disorder}. Subsequently, transport coefficients were calculated via the standard linear response theory. More details of the numerical simulation can be found in the Methods section.

Figure \ref{fig:2}c,d shows the electrical conductivity calculated from the CPA Green's function via the Kubo-Greenwood formalism.
With an impurity concentration $c$ ranging from $5\times10^{-4}$ to $0.5$, the Fermi level dependence of conductivity gets weaker and the magnitude of conductivity reaches a minimum at around $c=0.006$. Further increasing the impurity concentration leads to an increase of conductivity at a given Fermi level position. Such anomalous increase of conductivity with impurity concentration is difficult to understand based on a single surface model \cite{thy_prb,2d_transport}, which suggests the essential role of the bulk of a 3D TI in surface conduction.
In the following, we reveal that the anomalous increase of conductivity is a signature of a different type of spin dynamics and manifests a new spin generation mechanism in 3D TIs.

\begin{figure}[!hbp]
\begin{center}
\includegraphics[width=\textwidth]{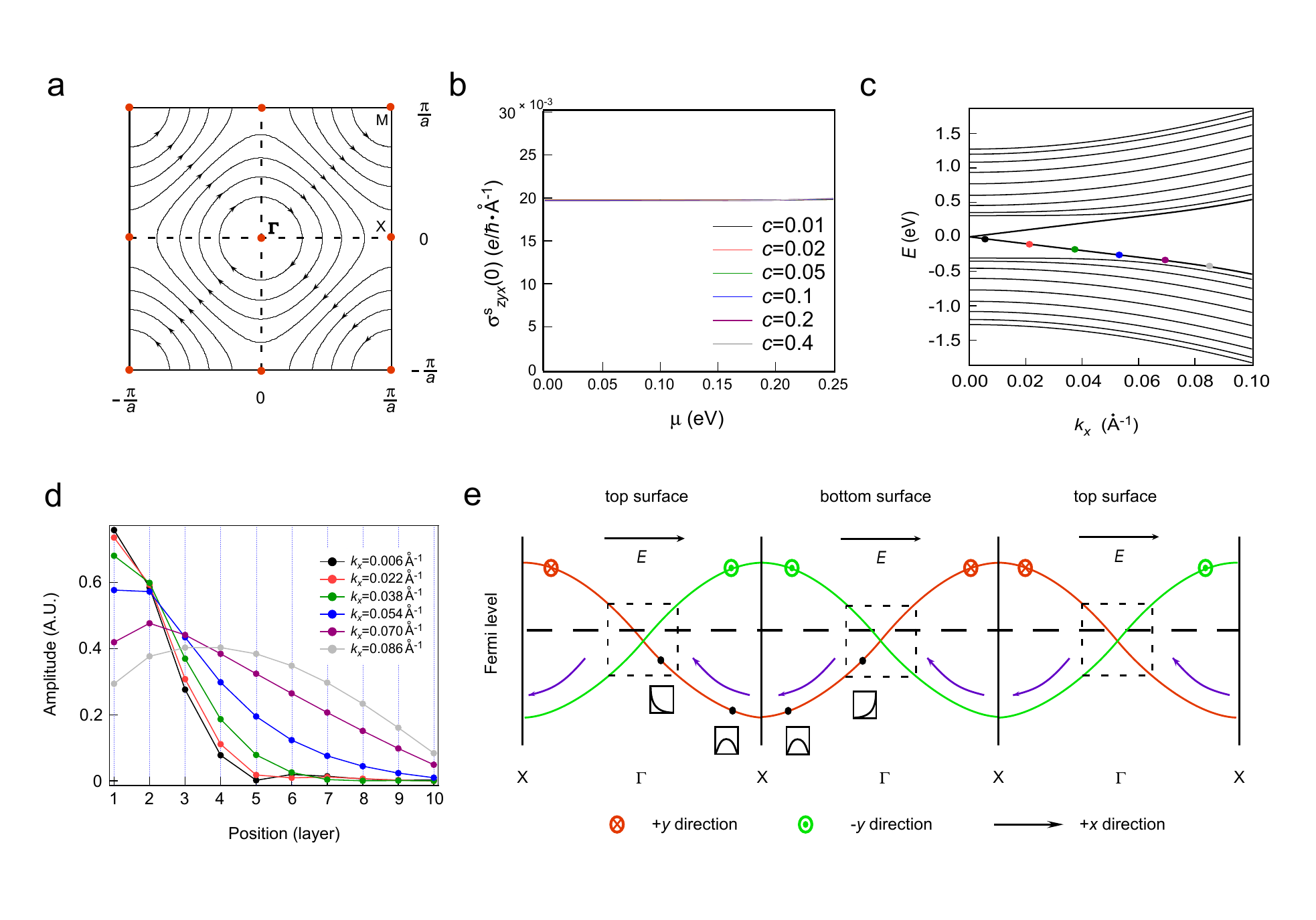}
\caption{The spin generation mechanism. {\bf a,} The chiral spin texture over the entire BZ for the conduction band of the top surface. The spin orientations on TRIM points (red dots) are degenerate due to the Kramers theorem. {\bf b,}  The DC spin Hall conductivity $\sigma^s_{zyx}(0)$ obtained from numerical simulation plotted against the Fermi level position $\mu$ at different impurity concentrations $c$. It is seen that the bulk spin current is independent of both Fermi level position and impurity concentration. {\bf c,} Energy dispersion for the top surface near the $\Gamma$ point along $\Gamma-\mathrm{X}$ direction. The spin polarization of the colored dots is in the $+y$ direction. {\bf d,} The evolution of electronic wave functions along the colored dots in {\bf c}.  As the magnitude of wave vector becomes larger, the wave function of an electron gradually evolves from being localized near the surface to extensive in the bulk. {\bf e,} A schematic plot for the spin generation mechanism in the extended BZ view. The drift motion along $x$ direction in $k$ space gives rise to spin transfer in the $z$ direction of the real space, which results in a pure spin current through the bulk. The red/green arrows pointing into/out of the page indicate spin polarization. The purple arrows indicate the direction of drift motion of electrons under an electric field in the $+x$ direction. The dashed horizontal line indicates the Fermi level position. The schematic drawing under the black dots denote the electronic wave functions. The dashed boxes denote the true surface state regions. It is essential that the Fermi level lies within the box regions for the spin transfer mechanism to apply. }\label{fig:3}
\end{center}
\end{figure}

To start discussions on spin dynamics, we notice that spin is not a predefined quantity in Hamiltonian (\ref{eq:1}).
Although common 3D TIs such as $\mathrm{Bi}_2\mathrm{Se}_3$ are known to have chiral spin texture on the surface states, the spin polarization is not $100\%$ \cite{spin_polarization}. Nevertheless, one can always talk about a ``pseudo-spin'' which is defined to exactly match the energy eigenstates and has all essential features of the real spin \cite{TI_book}. Here we take the definition
\begin{eqnarray}
S_x&=&-\ui\alpha_y\alpha_z\beta\nonumber\\
S_y&=&-\ui\alpha_z\alpha_x\beta\\
S_z&=&\ui\alpha_x\alpha_y\nonumber
\end{eqnarray}
It can be verified that such definition satisfies all symmetry requirements of the real spin and shows a chiral spin texture near the $\Gamma$ point, as shown in Figure \ref{fig:3}a. Note that a unique spin polarization can be specified for all points in the $\b{k}$-space except for those time reversal invariant momenta (TRIM) where the Kramers theorem asserts the degeneracy of the two opposite spin polarizations.

With the above definition of spins, the bulk spin current density $j^{\mathrm{s}}_{zy}$ denoting the transport of $y$-spins in the $z$-direction can be defined as
\[
j^{\mathrm{s}}_{zy}=(P_y^+v_zP_y^+-P_y^-v_zP_y^-)/\Omega_3
\]
where $P_y^+$ and $P_y^-$ are spin projection operators in the $+y$ and $-y$ directions, $v_z$ is the velocity operator in the $z$ direction, and $\Omega_3$ is the volume of the slab serving as a normalization factor. With the above expression, the spin Hall conductivity $\sigma^{\mathrm{s}}_{zyx}$ defined by
\[
j^{\mathrm{s}}_{zy}=\sigma^{\mathrm{s}}_{zyx}E_x
\]
can be calculated via the standard linear response theory. Due to the even symmetry of spin current $\b{j}^s$ under time reversal $\mathscr{T}$, the resultant expression for the spin Hall conductivity is different from that for the electrical conductivity (Kubo-Greenwood formula), but contains a term which involves all states below the Fermi level. This term has been  thoroughly reviewed in \cite{AHE} for the calculation of electrical conductivity in a $\mathscr{T}$-symmetry broken system and also discussed in a recently published article \cite{SHE} for the calculation of spin Hall conductivity. The emergence of this term in our system indicates the non-dissipative nature of the spin current, which has already been demonstrated possible for a wide class of traditional semiconductors \cite{SHI_science,SHI_prb}. We leave the details of derivation to the Supplementary Information and plot the calculated spin Hall conductivity $\sigma^{\mathrm{s}}_{zyx}$ in Figure \ref{fig:3}b. It is clear that the magnitude of the spin Hall conductivity is independent of both the Fermi level position (must be within the bulk bandgap) and the surface impurity concentration.

Although the existence of a bulk spin current in 3D TI has been predicted analytically through topological argument \cite{QSHE_TI}, a visualization of the spin transfer mechanism is not yet available so far. Neither has its relevance to the transport behavior of the gapless surface states been studied ever. In the following, we present an intuitive picture of the spin transfer in a 3D TI slab and uncover its close relationship with the surface-bulk correspondence of a 3D TI.

We notice that despite the chiral spin texture over the entire Brillouin zone (BZ), only states with small magnitude of momentum are truly localized on the surface. Figure \ref{fig:3}c and d show the evolution of electronic wave functions as the wave vector $\b{k}$ approaches the BZ boundary from the $\Gamma$ point. It is seen that, beyond a certain point, electronic wave functions become extended through the entire bulk and the surface band has essentially merged into bulk bands. States beyond this merging point should be classified as bulk states although they lie on the same branch of energy sub-band as true surface states.

Imagine applying a weak electric field to this system in the $+x$ direction and examine the $\Gamma-\mathrm{X}$ line in the extended BZ view. Due to the inversion symmetry in the $z$ direction, all bands are doubly degenerate. We notice however, in order for the spin texture to be continuous, every top surface branch must be connected to the adjacent bottom surface branch and vice versa, as shown in Figure \ref{fig:3}e. This alternating structure exists across all TRIM points in our system, and is distinctively different from a normal band which smoothly connects to itself at the BZ boundary. Consider an electron on the bottom surface with its spin polarized in $+y$ direction. Under the driving of the electric field, this electronic state drifts to $-x$ direction in $k$ space and merges into the bulk valance band. Upon further drifting, this electron finally enters the top surface with its spin in $+y$ direction unchanged. Simultaneously, an electron with spin polarized in the $-y$ direction will drift from the top surface to bottom surface. The drift motion across the $\mathrm{X}$ point is similar to the Klein tunneling of Dirac Fermions in the sense that, in order for a certain spin to be continuous, the electron must tunnel to another band rather than return to its original band. Overall, each of these processes corresponds to a unit spin-pair exchange between the bottom surface and the top surface.  Thus a longitudinal electric field induces a transverse pure spin current through the bulk, which plays the role of a spin injector for the two surfaces, as described in Figure \ref{fig:1}a. During this process, it is essential that the Fermi level lies within the gap, because there exists another pair of merging points near the conduction band edge. If the Fermi level is above these points as well, there would be an opposite process which leads to the cancelation of net spin current. This is of course consistent because the system in this case is not a TI any more.

\begin{figure}[!hbp]
\begin{center}
\includegraphics[width=\textwidth]{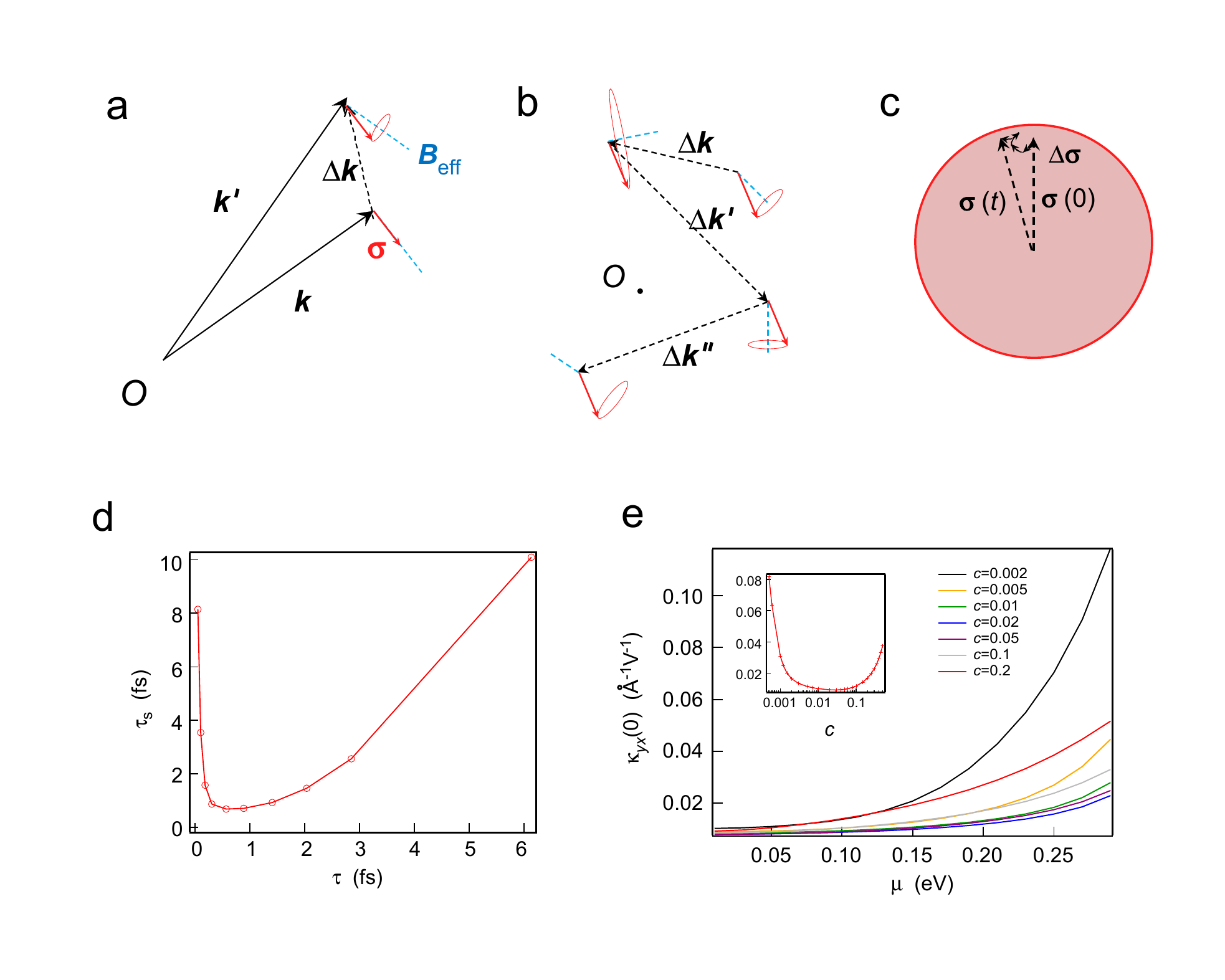}
\caption{Spin relaxation mechanism under high level of nonmagnetic disorder. {\bf a,} When scatterings are frequent, an  electron suffers from rapid momentum change where the electronic spin does not have time to follow the instantaneous energy eigenstate given by adiabatic perturbation theory. Instead, the electronic spin precesses about the instantaneous energy eigenstate which serves as an effective magnetic field $\b{B}_{\mathrm{eff}}$. {\bf b,} Frequent scatterings constantly change the precession axis of spin. During the interval of two consecutive scatterings the spin can only precess for a small angle. {\bf c,} The spin ends up doing a random walk on a unit sphere. The more frequent the scatterings are, the less efficient this random walk is, and consequently spin can preserve its original direction for a longer time. {\bf d,} The numerical simulation result for spin relaxation time $\tau_{\mathrm{s}}$ and momentum relaxation time $\tau$ at Fermi level $\mu=0.13\ \mathrm{eV}$. As expected, when disorder is high, $\tau_{\mathrm{s}}$ and $\tau$ have an inverse dependence as in the traditional Dyakonov-Perel spin relaxation mechanism. {\bf e,} The simulated DC electro-spin susceptibility $\kappa_{yx}(0)$ plotted against the Fermi level $\mu$ at different impurity concentrations. The inset is $\kappa_{yx}(0)$ at a fixed Fermi level $\mu=0.13\ \mathrm{eV}$ versus impurity concentration $c$. It is clear that under high disorder, the accumulated spin density increases with impurity concentration. }\label{fig:4}
\end{center}
\end{figure}

Unlike charge, spin is not a conserved quantity in our system. The spins injected onto the surface suffer from immediate relaxation. The scenario is slightly different from both the Hall effect and the 2D spin Hall effect. In 3D the spin relaxation is actually necessary for the system to reach a steady state, as detailed in Supplementary Note 1. The ultimate spin density accumulated on the surface is determined by the spin relaxation time $\tau_{\mathrm{s}}$. In the following, we provide an intuitive physical picture of the spin relaxation dynamics under the eigenbasis defined by $H_0$. More details of this picture can be found in Supplementary Note 2. This is essentially an interaction picture which splits the Hamiltonian into a free part $H_0$ and an interaction part $U$.
Due to the momentum-spin locking, each electron senses an effective magnetic field $\b{B}_{\mathrm{eff}}$ according to its wave vector $\b{k}$. When an electron with its spin aligned with $\b{B}_{\mathrm{eff}}$ suffers from a momentum change $\hbar\Delta \b{k}$ due to the scattering of an impurity potential, its spin may no longer align with the new $\b{B}_{\mathrm{eff}}$. If scatterings are rare, i.e. the time it takes for momentum to change by a unit amount is long, the adiabatic perturbation theory predicts that the new spin must evolve to the new energy eigenstate, i.e. rotate to the direction of the new $\b{B}_{\mathrm{eff}}$. If scatterings are frequent, however, the spin does not have time to follow $\b{B}_{\mathrm{eff}}$ and will precess about the instantaneous $\b{B}_{\mathrm{eff}}$, as shown in Figure \ref{fig:4}a. Frequent scatterings constantly change the precession axis and the spin ends up doing a random walk on a unit sphere, as shown in Figure \ref{fig:4}b and c. The more frequent momentum scattering is, the less effective the random walk is, and the spin will preserve its original direction for a longer time. Therefore, the spin relaxation time $\tau_{\mathrm{s}}$ inversely depends on the momentum relaxation time $\tau$, just like in the traditional Dyakonov-Perel mechanism \cite{DP}. One point to note is that, since disorder is only present on the surface, states outside the dashed box of Figure \ref{fig:3}e are unaffected by scattering and the previously discussed spin transfer mechanism remains valid even under strong surface disorder.

If the above physical picture is correct, the ultimate spin density accumulated on the surface should increase with the increase of disorder. We calculated the spin relaxation time $\tau_{\mathrm{s}}$ and electro-spin susceptibility $\kappa_{yx}$ via standard linear response theory. The results are shown in Figure \ref{fig:4}d and e, which perfectly agree with the expectation. Combined with the fact that velocity operator is proportional to spin on the surface, it is not difficult to understand the anomalous increase of conductivity as well.

\begin{figure}[!hbp]
\begin{center}
\includegraphics[width=\textwidth]{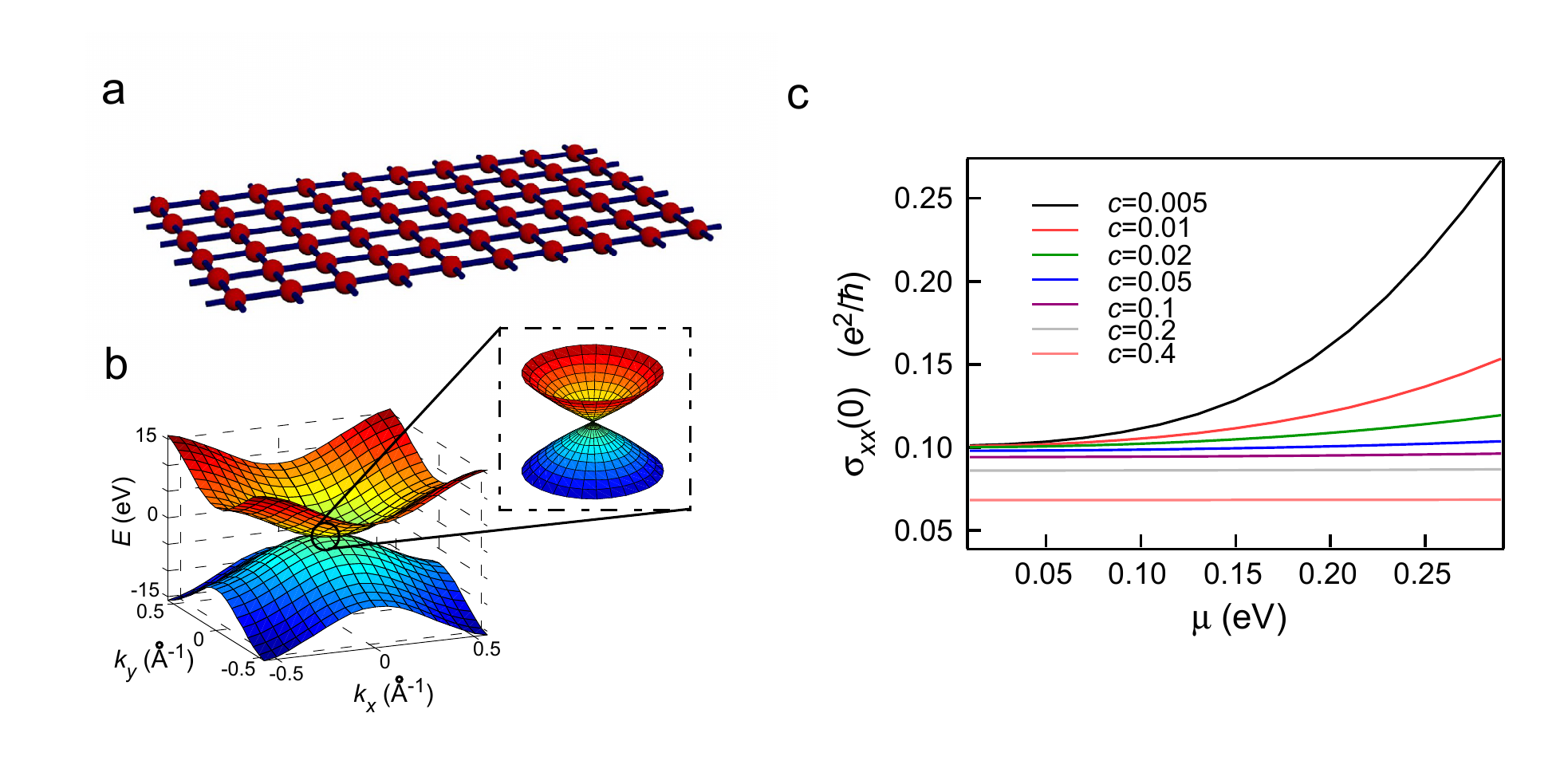}
\caption{Simulation for a single layer with the same model to illustrate the importance of spin dynamics in our previous system. {\bf a,} The geometry of the system. Each atom on this layer has a probability of $c$ to be occupied by a vacancy and $(1-c)$ to be intact. The on-site energy of a vacancy is brought to infinity. Only nearest neighbor hopping is allowed. {\bf b,} The energy dispersion of this single layer system. Simulation parameters are identical to before except for the mass term $\Delta$ is set to zero in order to have a Dirac cone near the $\Gamma$ point. The band structure looks almost identical to the surface bands of the 3D TI slab discussed before. {\bf c,} The DC conductivity $\sigma_{xx}(0)$ of this single layer versus Fermi level at different impurity concentrations. Despite the similarity in band structures, this single layer does not have anomalous increase of conductivity at high disorder, in contrast with our previous system. }\label{fig:5}
\end{center}
\end{figure}

It is worth noting that, in contrast to a common belief, the behavior of conductivity of a Dirac system under high disorder is not simply governed by the dispersion relationship. The anomalous increase of conductivity is closely related to the spin generation and relaxation mechanism. To illustrate this point, we calculated the conductivity of a single layer of atoms within the same model (Figure \ref{fig:5}a). By setting the mass term $\Delta=0$, the band structure of this 2D system has almost identical shape as the previous surface state, as shown in Figure \ref{fig:5}b, but the spin generation mechanism discussed above is obviously absent.  With the increase of the impurity concentration, the conductivity of such system monotonically decreases towards zero, even in the high disorder range. Similar behavior has also been shown in graphene \cite{gra1,gra2}, where no anomalous increase of conductivity was found.

The spin dynamics in a 3D TI under strong surface disorder makes delicate connection with the scenario of a 2D TI. In fact, it is straightforward to apply the previously discussed spin transfer mechanism to a 2D TI, and obtain the universal quantized spin Hall conductivity $\sigma^{\mathrm{s}}_{\mathrm{2D}}=e/(\pi\hbar)$. Different from the 3D case, in the edge channel of a 2D TI, the spin (understood as pseudo-spin when necessary) is a conserved quantity which does not relax. Consequently, the electro-spin susceptibility is infinite, which means no external field is needed to support the edge spin accumulation and charge current. The quantized and finite channel conductivity $e^2/h$ is actually a contact effect while the channel itself is dissipationless \cite{HgTe_thy}. On the surface of a 3D TI, however, spin is not conserved due to an additional angular degree of freedom of the wave vector $\b{k}$. The electro-spin susceptibility is thus finite and transport is dissipative. Strong surface disorder greatly suppresses spin relaxation and brings the system closer to the situation of a 2D TI, leading to a more efficient spin generation. In this view, strong surface disorder can be beneficial for spintronic devices, in contrast with the common belief. Technically, it is obviously of more convenience to induce high disorder on a surface than to make it pure and pristine.

Experimentally, the anomalous increase of conductivity in 3D TIs has already been hinted by results from several groups, yet researchers do not generally regard it as an intrinsic property of the TI surface. Field effect measurements in TIs have often shown a high minimum conductivity even when the Fermi level is tuned to the charge neutral point \cite{tran_4,tran_5,tran_6,tran_7,tran_8}, which is much larger than expected by normal transport theory assuming low disorder \cite{thy_prb,2d_transport}. Although they are often attributed to bulk conduction \cite{tran_5} or surface electron puddle formation \cite{tran_6,trans_gra_thy}, a closer look at these models reveals several problems, which are discussed in details in Supplementary Note 3. Theory presented in this paper, however, provides a simple and natural way to understand these observations.  Recently, it has been directly observed in exfoliated $\mathrm{BiSbTeSe}_2$ nanoflakes that after argon ion milling treatment to create more surface defects, the sample becomes more conductive \cite{Neg_MR}, although this effect was not understood.

Our simulation results also suggest that under strong surface disorder, with the magnitude of transport coefficients increased, their Fermi level sensitivity has dropped, which is also beneficial for the design of 3D TI-based spintronic devices. This is because in ambient environment, the Fermi level on the surface is subject to unintentional change due to contamination and degradation \cite{surf_degrad,surf_degrad2}. Sensitive Fermi level dependence renders the device less stable and robust in air.

In summary, our CPA simulation on a 4-band lattice model reveals a new and efficient spin generation mechanism in a 3D TI. The topological nature of band structures demands a pure spin current between the two opposite surfaces and consequent spin transfer. High level of nonmagnetic disorder can suppress spin relaxation, allowing substantial increase of spin accumulation at the surfaces. Such mechanism is manifested by the anomalous increase of conductivity under high disorder. This work not only provides valuable fundamental physical insights on spin/charge transport in 3D TIs, but also offers important guidance to the design of 3D TI-based spintronic devices.

\section*{Methods}
The numerical calculation results were obtained via CPA and the standard linear response theory. Details of the calculation are given below. In the following, we use convention $\hbar=1$.
\subsection*{CPA and Self-energy}
The Hamiltonian of a clean lattice in the main text can be written in a block diagonal form
\[
H_0(k_x,k_y)=\left(
\begin{array}{cccc}
\epsilon(k_x,k_y) & -\ui\frac{A_z}{2}\alpha_z-B_z\beta & 0 & \ldots\\
\ui\frac{A_z}{2}\alpha_z-B_z\beta & \epsilon(k_x,k_y) & -\ui\frac{A_z}{2}\alpha_z-B_z\beta &0\\
0& \ui\frac{A_z}{2}\alpha_z-B_z\beta & \epsilon(k_x,k_y) & \\
\vdots & 0 & &\ddots
\end{array}
\right)
\]
where each element in this matrix is a $4\times4$ matrix. The $N$ rows and columns denote the $N$ layers of the slab. The on-site energy $\epsilon(k_x,k_y)$ takes the form
\[
\epsilon(k_x,k_y)=A(\sin k_x a \alpha_x+\sin k_y a \alpha_y)+\left[\Delta-2B_z-4B\left(\sin^2\frac{k_xa}{2}+\sin^2\frac{k_ya}{2}\right)\right]\beta
\]
Overall, the Hamiltonian is $4N\times4N$. Diagonalizing this matrix gives the band structure of a clean lattice without impurities.

We consider impurities of an on-site scalar potential $U$ on the top and bottom layers, which takes the form of
\[
U=\mathrm{diag}[u,u,u,u,0,0,\ldots,0,0,u,u,u,u]
\]
Each site has a probability of $c$ subject to potential $U$ and probability $(1-c)$ subject to potential $0$, which makes the entire system essentially a binary alloy. Such configuration assumes non-physical correlation between the appearance of an impurity on the top and bottom layers. Yet from practical consideration, for a sufficiently thick slab, the crosstalk between the top and bottom layers should vanish, which justifies the binary alloy model of the above.

In CPA, the configurationally averaged impurity potential is denoted by a $\b{k}$-independent self-energy $\Sigma(\omega)$. In the binary alloy case, $\Sigma(\omega)$ is determined by the iterative equation
\[
\Sigma(\omega)=cU[1-G(\omega)(U-\Sigma(\omega))]^{-1}
\]
where $G(\omega)$ is the on-site Green's function
\[
G(\omega)=\sum_{\b{k}}[\omega-H_0(\b{k})-\Sigma(\omega)]^{-1}
\]
Due to the symmetry of this problem, $\Sigma(\omega)$ is actually a scalar on the top and bottom layers only. The real and imaginary parts of the self-energy are plotted in Supplementary Figure 1.

\subsection*{Conductivity}
The velocity operator $\b{v}$ takes the form
\begin{eqnarray}
 \b{v}&=&\frac{\partial H}{\partial \b{k}}\\
 &=&\left(
 \begin{array}{cccc}
 \frac{\partial \epsilon}{\partial \b{k}} & & &\\
 & \frac{\partial \epsilon}{\partial \b{k}}& &\\
 & & \frac{\partial \epsilon}{\partial \b{k}} &\\
 & & & \ddots
 \end{array}
 \right)
\end{eqnarray}
According to the Kubo formula, the DC conductivity at $T=0$ is
\[
\sigma_{xx}(0)=\frac{e^2}{4\pi^3}\int \mathrm{Tr}[v_x(\b{k})\mathrm{Im} G(\b{k},\mu)v_x(\b{k})\mathrm{Im} G(\b{k},\mu)] \ud^2\b{k}
\]
The conductivity of the top surface is half of this value
\[
\sigma_{xx}^{\mathrm{top}}(0)=\frac{1}{2}\sigma_{xx}(0)
\]

\subsection*{Definition of spin operators and the chiral spin texture}
Spin is not a predefined quantity in the original Hamiltonian. Unless we are interested in a direct coupling to an external magnetic field, the four state basis vectors can be thought to have arbitrary spin polarizations, which is essentially a ``pseudo-spin''. Even in real topological insulators such as $\mathrm{Bi}_2\mathrm{Se}_3$, the pseudo-spin does not exactly match the real spin. Nevertheless, the definition of a ``pseudo-spin'' $\b{S}$ must satisfy a couple of restrictions:
(1) $\b{S}$ is an Hermitian operator. $S_i^\dag=S_i$.
(2) The components of $\b{S}$ satisfy the anti-commutation rules. $S_iS_j=\delta_{ij}+\ui\epsilon_{ijk}S_k$.
(3) $\b{S}$ is a pseudo-vector. It transforms like a vector under in-plane (xy) rotation but does not flip sign under space inversion. $\beta S_i \beta=S_i$.
(4) In order to comply with the chiral surface spin texture, we require $\b{S}$ be polarized along y-direction when $k_y=0$. Thus $[S_y,H(k_x,0)]=0$.

Note that we do not include $\hbar/2$ in our definition, so this pseudo-spin has dimension 1 instead of angular momentum.
It can be easily verified that the following expressions are a good representation of ``spin'' in our system.
\begin{eqnarray}
S_x&=&-\ui\alpha_y\alpha_z\beta\\
S_y&=&-\ui\alpha_z\alpha_x\beta\\
S_z&=&\ui\alpha_x\alpha_y
\end{eqnarray}
The spin in an arbitrary direction $\theta$ within the $xy$ plane is thus
\[
S(\theta)=S_x\cos\theta+S_y\sin\theta
\]
If there exists an angle $\theta$ for an arbitrary point in $k$-space $(k_x,k_y)$ such that
\[
[S(\theta),H_0(k_x,k_y)]=0
\]
then $S(\theta)$ and $H_0(k_x,k_y)$ have common eigenstates and the energy eigenstates can be assigned a unique spin polarization. It is not difficult to obtain
\[
\tan\theta=-\frac{\sin k_xa}{\sin k_ya}
\]

When the magnitude of $\b{k}$ is small, we recover to the well-known chiral spin texture
\[
\tan\theta\approx-\frac{k_x}{k_y}
\]
while when $\b{k}$ gets large, the spin texture deforms to adapt to the tetragonal symmetry of the BZ, as shown in the main text.

\subsection*{Surface spin density and electro-spin susceptibility}
The spin density accumulated on the top surface is
\begin{eqnarray}
s_y^{\mathrm{top}}&=&\frac{1}{\Omega}
\left(
 \begin{array}{ccc:ccc}
S_y & & & & &\\
 & S_y& & & &\\
 & & \ddots & & &\\
 \hdashline
 & & & 0 & &\\
 & & & &0&\\
 & & & & & \ddots
 \end{array}
 \right)
\end{eqnarray}
At $T=0$ the electro-spin susceptibility is calculated similar to the electric conductivity
\begin{eqnarray}
 \kappa_{yx}(0)&=&\frac{e}{\pi}\Tr[s_y^{\mathrm{top}}\Im G(\mu)v_{\alpha}\Im G(\mu)]\\
 &=&\frac{e\Omega}{4\pi^3}\int \ud^2 \b{k}\Tr\left[s_y^{\mathrm{top}}\Im G(\b{k},\mu)v_{\alpha}(\b{k})\Im G(\b{k},\mu)\right]
\end{eqnarray}
Note that the definition of $s_y^{\mathrm{top}}$ contains a factor of $1/\Omega$ and thus the above expression is actually independent of the box size $\Omega$.

\subsection*{Bulk spin current and spin Hall conductivity}
The z-position operator is
 \[
 z=a_z\cdot\mathrm{diag}[-1,-1,-1,-1,-2,-2,-2,-2,-3,-3,-3,-3,\ldots]
 \]
 Thus the z-velocity operator
 \begin{eqnarray}
 v_z&=&-\ui[z,H]\\
 &=&a_z\left(
 \begin{array}{cccccc}
 0  &  -\frac{A_z}{2}\alpha_z+\ui B_z\beta & & & &\\
 -\frac{A_z}{2}\alpha_z-\ui B_z\beta & 0  & -\frac{A_z}{2}\alpha_z+\ui B_z\beta & & &\\
  & -\frac{A_z}{2}\alpha_z-\ui B_z\beta & 0 & \ddots & & \\
   & & \ddots & \ddots &  &
 \end{array}
 \right)
 \end{eqnarray}
 Since we are interested in the flux across a certain intermediate layer, we restrict the velocity operator to be only between two adjacent layers in the middle
 \begin{eqnarray}
 v_z^m&=&a_z\left(
 \begin{array}{cccccc}
   &   & & & &\\
     &   & & & &\\
  & 0  & -\frac{A_z}{2}\alpha_z+\ui B_z\beta & & &\\
  & -\frac{A_z}{2}\alpha_z-\ui B_z\beta & 0 &  & & \\
   & &  &  &  &\\
     &   & & & &
 \end{array}
 \right)\\
 \end{eqnarray}
 In order to discuss the spin current, we need to define a spin projection operator
 \begin{eqnarray}
 P_y^+&=&\mathbb{I}\otimes|S_y=+1\rangle\langle S_y=+1|\\
&=&\mathbb{I}\otimes\frac{1}{2}\left(
\begin{array}{cccc}
1 & \ui &  &  \\
-\ui & 1 &  &  \\
  &  & 1 & -\ui\\
  &  & \ui & 1
  \end{array}
  \right)\\
  &=&\mathbb{I}\otimes\frac{1}{2}(1+S_y)
  \end{eqnarray}
  \begin{eqnarray}
 P_y^-&=&\mathbb{I}\otimes|S_y=-1\rangle\langle S_y=-1|\\
&=&\mathbb{I}\otimes\frac{1}{2}\left(
\begin{array}{cccc}
1 & -\ui &  &  \\
\ui & 1 &  &  \\
  &  & 1 & \ui\\
  &  & -\ui & 1
  \end{array}
  \right)\\
  &=&\mathbb{I}\otimes\frac{1}{2}(1-S_y)
  \end{eqnarray}
The spin current operator is then
\begin{eqnarray}
j^s_{zy}&=&(P_y^+v^m_zP_y^+-P_y^-v^m_zP_y^-)/(a_z\Omega)\\
&=&\frac{1}{\Omega}\left(
 \begin{array}{cccccc}
   &   & & & &\\
     &   & & & &\\
  & 0  & -\ui\frac{A_z}{2}\alpha_x\beta- B_z\alpha_z\alpha_x & & &\\
  & -\ui\frac{A_z}{2}\alpha_x\beta+ B_z\alpha_z\alpha_x & 0 &  & & \\
   & &  &  &  &\\
     &   & & & &
 \end{array}
 \right)
\end{eqnarray}
Under time reversal
\[
\mathscr{T}(j^s_{zy})=j^s_{zy}
\]
which is different from charge current or spin density. Thus the spin Hall conductivity is not only determined by the Green's function at the Fermi level but contains an additional term contributed by all states occupied \cite{AHE,SHE}.
\begin{eqnarray}
\sigma^s_{zyx}(0)&=&\frac{e}{4\pi}\Tr [j^s_{zy}G(\mu)v_x G^\dag(\mu)-j^s_{zy}G^\dag(\mu)v_x G(\mu)]\\
&&+\frac{e}{4\pi}\int_{-\infty}^{\mu}\ud \lambda\Tr\left[-j^s_{zy}\frac{\ud G(\lambda)}{\ud \lambda}v_x G(\lambda)+j^s_{zy}G(\lambda)v_x\frac{\ud G(\lambda)}{\ud \lambda}+\mathrm{h.c.}\right]\\
&=&\frac{e\Omega}{16\pi^3} \int\ud^2\b{k}\Tr [j^s_{zy}G(\b{k},\mu)v_x(\b{k}) G^\dag(\b{k},\mu)-j^s_{zy}G^\dag(\b{k},\mu)v_x(\b{k}) G(\b{k},\mu)]\\
&&+\frac{e\Omega}{16\pi^3}\int_{-\infty}^{\mu}\ud \lambda\int\ud^2\b{k}\Tr\left[-j^s_{zy}\frac{\ud G(\b{k},\lambda)}{\ud \lambda}v_x(\b{k}) G(\b{k},\lambda)+j^s_{zy}G(\b{k},\lambda)v_x(\b{k})\frac{\ud G(\b{k},\lambda)}{\ud \lambda}
+\mathrm{h.c.}\right]
\end{eqnarray}
Again due to the $1/\Omega$ factor in the definition of $j^s_{zy}$, the above expression is independent of the box size $\Omega$.

\subsection*{Spin relaxation time}
Imagine applying a pulse electric field to our system
\[
E(t)=E_0t_0\delta(t)
\]
Empirically the induced spin has the asymptotic form
\[
s(t)=s_0\ue^{-\frac{t}{\tau_s}}\theta(t)
\]
where the function $\theta(t)=0$ when $t<0$ and $\theta(t)\rightarrow 1$ as $t\rightarrow +\infty$. Consider the Fourier transform
\begin{eqnarray}
\hat{s}(\omega)&=&\int_{-\infty}^{+\infty} s(t)\ue^{\ui\omega t}\ud t\\
&=&s_0\int_{-\infty}^{+\infty}\theta(t)\ue^{\left(\ui\omega-\frac{1}{\tau_s}\right) t}\ud t
\end{eqnarray}
For sufficiently small $\omega$, the details of the rising part of $s(t)$ characterized by $\theta(t)$ becomes unimportant, thus we replace $\theta(t)$ with the step function $\Theta(t)$ and obtain
\begin{eqnarray}
\hat{s}(\omega)&=&s_0\int_{0}^{+\infty}\ue^{\left(\ui\omega-\frac{1}{\tau_s}\right) t}\ud t\\
&=&-\frac{s_0}{\ui \omega-\frac{1}{\tau_s}}
\end{eqnarray}
The Fourier transform of the electric field is just a constant
\[
\hat{E}(\omega)=E_0t_0
\]
Thus the electro-spin susceptibility
\[
\kappa(\omega)\propto\frac{1}{\ui \omega-\frac{1}{\tau_s}}
\]
and the spin relaxation time can be extracted as
\begin{eqnarray}
\tau_s&=&-\ui\frac{1}{\kappa}\left.\frac{\ud \kappa(\omega)}{\ud \omega}\right|_{\omega=0}\\
&=&-\frac{1}{2}\frac{\chi''(0)}{\kappa(0)}\label{eq:2}
\end{eqnarray}
It is necessary to point out that the spin relaxation process can be thought as an eigen mode with a complex frequency $\omega^\ast$ on the lower half plane. $\omega^\ast$ is a pole of the response function $\chi(\omega^\ast)=\infty$. The spin relaxation time is determined by the imaginary part of the pole closest to the real axis
\[
\tau_s=-\frac{1}{\Im (\omega^\ast)}
\]
The expression (\ref{eq:2}) is based on the low frequency expansion of $\chi(\omega)$ which may not give the exact pole position. \cite{srelax} has shown that for a perfect Dirac-cone dispersion the spin relaxation time found by the exact pole is twice as the value found by low frequency expansion. Nevertheless, apart from an order 1 factor, low frequency expansion should give a reasonable estimate of the true spin relaxation time.

\vspace{0.5cm}

\textbf{Acknowledgement}

This work was supported by the U.S. National Science Foundation Grant DMR-1310678 (D.Y.), DMR-1306048 (R.R.P.S.), DMR-1411336 (S.Y.S.) and 2015-2016 UC Davis Summer Graduate Student Researcher Award (X.P.).
The authors thank Prof. N. Curro, Prof. R. Scalettar and Prof. C-Y Fong for extensive discussions. X.P. thanks Xi Chen for the assistance in the preparation of figures.

\vspace{0.5cm}
\textbf{Author contributions}

X.P. built up the model and performed the numerical simulation. S.Y.S. offered assistance in simulation code construction.  All authors participated in scientific discussion and manuscript preparation.

\vspace{0.5cm}
\textbf{Additional information}

Supplementary information is available in the online version of the paper. Correspondence and request for materials should be made to D.Y.

\vspace{0.5cm}
\textbf{Competing financial interest}

The authors declare no competing financial interest.


\end{document}


%
%

\section*{Supplementary Figures}
\begin{figure}[!hbp]
\begin{center}
\includegraphics[width=\textwidth]{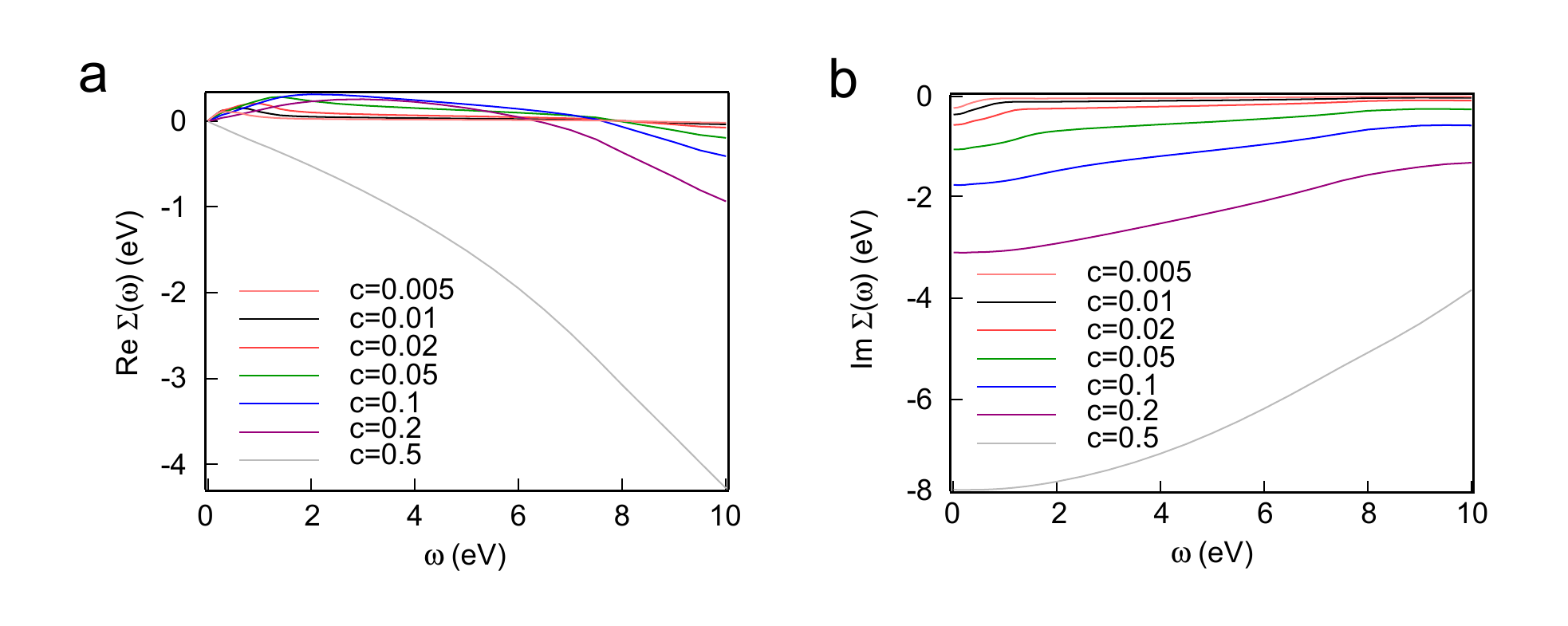}
\caption{The self-energy $\Sigma(\omega)$ computed for the 3D TI slab as discussed in the Methods section of the main text. {\bf a,} The real part and {\bf b,} The imaginary part.}
\end{center}
\end{figure}

\section*{Supplementary Note 1: Spin non-conservation and the role of spin relaxation in the spin Hall effect}
As pointed out in the main text, spin is not a conserved quantity in this system. However, the non-conservation of spin does not completely break off the connection between spin density and spin current density. It only requires that a spin relaxation term be added to the continuity equation and taken with care. The bulk of our system is disorder-free where time reversal symmetry prohibits any spin accumulation at a steady state, thus spin accumulation and relaxation can only happen on the surface.
An empirical equation for surface spin density $s$ and bulk spin current $j^s$ can be written down as
\[
\frac{\ud s}{\ud t}=j^s-R(s)
\]
where $R(s)$ is the surface spin relaxation rate which can be roughly expressed as $s/\tau_s$. At a steady state, one has
\[
j^s=R(s)\approx\frac{s}{\tau_s}
\]
from which the accumulated surface spin density $s$ is determined. If there were no such spin flip term, the system would never reach a steady state.

The situation for the spin Hall effect is subtly different from the Hall effect. In the Hall effect, the accumulated surface charge causes a lateral voltage drop to resist further accumulation of charge, such that at a steady state the lateral current $j\equiv 0$. In the spin Hall effect, however, the spin accumulation does not lead to any counter force for the spin current, and there is a persistent bulk spin current as long as the longitudinal electric field exists. This term has to be canceled by a spin flip term to reach the steady state.

The situation is also different from the 2D case. In the main text, we have argued that for the steady state of a 2D quantum spin Hall system there is actually no voltage drop along the conduction channel, but only across the contact. Therefore the lateral spin current $j^s\equiv 0$ similar to the Hall effect case. Thus it's possible to define a conserved spin current for a 2D system, but not for a 3D system where the voltage drops across the system itself.

Among the literature of spin Hall effect, some \cite{SHI} did attempt to define a conserved spin component. Yet from the above argument it seems that in the spin Hall effect, a properly defined, physically observable ``spin'' which manifests itself as a surface spin accumulation should be nonconserved.

\section*{Supplementary Note 2: About the Dyakonov-Perel spin relaxation mechanism}
A very tricky question regarding the Dyakonov-Perel spin relaxation mechanism in our system is the lack of $\b{k}$ space. Under high disorder $\epsilon_F\tau\lesssim 1$, the impurity potential cannot be regarded as a perturbation thus wave vector $\b{k}$ becomes an ill-defined quantity. Talking about the spin precessional random walk in this situation seems an unjustified story.

However, the lack of $k$-space is only true when we treat $H_0+U$ as a whole. The spin random walk picture of D-P spin relaxation mechanism is actually an interaction picture which splits the Hamiltonian into a free part $H_0$ and an interaction part $U$. $H_0$ provides the energy eigenstate bases while $U$ accounts for the time evolution of the wave function. The interaction Hamiltonian $U$ does not have to be much smaller than $H_0$. Wave vector $\b{k}$ is perfectly defined for $H_0$, which justifies the language of spin random walk.

To be more specific, we compare the situation of our system to a traditional 2DEG with Rashba spin splitting which is known to exhibit D-P mechanism.  The Hamiltonian in this case is
\[
H=H_b(\b{k})+H_s+U
\]
where $H_b(\b{k})$ is a spin-independent band energy, $H_s$ is the spin splitting energy which can be expressed as $\hbar v_F \b{\sigma}\cdot\b{k}$, $U$ is the scattering term. Here $v_F$ is just a parameter with no meaning of ``Fermi velocity''. The criterion for D-P mechanism is $|H_b|\gg|U|\gg|H_s|$ or $H_b\gg \hbar/\tau \gg \hbar v_F k$. The latter part of this criterion simply means the spin splitting structure is completely blurred by scattering. While on a 3D TI surface the entire band Hamiltonian is just the spin splitting energy, it certainly means a complete destruction of the k-space. The only difference from the traditional case is the lack of $H_b$ term. However, we will argue in the following that $H_b$ is not essential to the D-P mechanism.

We investigate the evolution of the system during a time $t$. We divide $t$ into a lot of infinitesimal intervals
\[
t=\Delta t_1+\Delta t_2+...+\Delta t_N
\]
The time evolution operator correspondingly breaks into
\[
\ue^{-\ui\frac{H}{\hbar}t}=\ue^{-\ui\frac{H}{\hbar}\Delta t_1}\ue^{-\ui\frac{H}{\hbar}\Delta t_2}...\ue^{-\ui\frac{H}{\hbar}\Delta t_N}
\]
Each interval can be separated in terms of the three terms of $H$
\[
\ue^{-\ui\frac{H}{\hbar}\Delta t}=\ue^{-\ui\frac{H_b}{\hbar}\Delta t}\ue^{-\ui\frac{H_s}{\hbar}\Delta t}\ue^{-\ui\frac{U}{\hbar}\Delta t}
\]
Now we consider an initial state
\[
|\b{k}\ra\otimes|\b{\sigma}\ra
\]
with momentum $\b{k}$ and spin in the $\b{\sigma}$ direction ($|\b{\sigma}\ra$'s are actually the coherent states of spin which form an over complete set in the spin space).

Acting $\ue^{-\ui\frac{U}{\hbar}\Delta t}$ on $|\b{k}\ra\otimes|\b{\sigma}\ra$ will scatter it to a different $\b{k}'$ with the amplitude determined by $U_{\b{k}'\b{k}}$
\[
\ue^{-\ui\frac{U}{\hbar}\Delta t}|\b{k}\ra\otimes|\b{\sigma}\ra=\sum_{\b{k}'} \left(\delta_{\b{k}'\b{k}}-\ui\frac{U_{\b{k}'\b{k}}}{\hbar}\Delta t\right)|\b{k}'\ra\otimes|\b{\sigma}\ra
\]
but leaving the spin vector $\b{\sigma}$ unchanged.

Acting $\ue^{-\ui\frac{H_s}{\hbar}\Delta t}$ on $|\b{k}\ra\otimes|\b{\sigma}\ra$ will precess the spin vector about the axis $\b{k}$ by an angle
\[
\ue^{-\ui\frac{H_s}{\hbar}\Delta t}|\b{k}\ra\otimes|\b{\sigma}\ra=|\b{k}\ra\otimes|\b{\sigma}+2v_F\b{k}\Delta t\times\b{\sigma}\ra
\]
but leaving the momentum $\b{k}$ unchanged.

Acting $\ue^{-\ui\frac{H_b}{\hbar}\Delta t}$ on $|\b{k}\ra\otimes|\b{\sigma}\ra$ does not change anything but simply induces a phase factor
\[
\ue^{-\ui\frac{H_b}{\hbar}\Delta t}|\b{k}\ra\otimes|\b{\sigma}\ra=\ue^{-\ui\frac{H_b(\b{k})}{\hbar}\Delta t}|\b{k}\ra\otimes|\b{\sigma}\ra
\]

Now we may assign each time interval $\Delta t_i$ an available $\b{k}_i$ to form an integral path
\[\label{eq:2}
\b{k}_1\times\Delta t_1 \rightarrow \b{k}_2\times\Delta t_2 \rightarrow \b{k}_3\times\Delta t_3 \rightarrow ... \rightarrow
\b{k}_N\times\Delta t_N
\]
The final state is just a sum over all paths.

Consider two extreme cases: (1) $|H_s|\gg|U|$. In this case the spin vector precesses by an appreciable angle far before the momentum has an appreciable probability to get scattered to a different value. The scattering is essentially an adiabatic rotation of the spin vector to the new energy eigenstate. Hence spin relaxation and momentum relaxation are bound together and we have $\tau_s=\tau$. (2) $|H_s|\ll|U|$. In this case the momentum gains an appreciable probability to be scattered to a different value far before the spin vector precesses by an appreciable angle. This will result in the precessional random walk picture of D-P mechanism. We thus expect $\tau_s \sim 1/\tau$.

The $H_b$ term, however, will cause some restriction to the above picture through the additional phase factor $\ue^{-\ui\frac{H_b(\b{k})}{\hbar}\Delta t}$. Consider a virtual variation to the path (\ref{eq:2}): we slightly change the lengths of $\Delta t_i$ and $\Delta t_{i+1}$ to $\Delta t_i+\delta$ and $\Delta t_{i+1}-\delta$. Since we have assumed $|H_b|\gg|H_s|$ and $|H_b|\gg|U|$, if $H_b(\b{k}_i)\neq H_b(\b{k}_{i+1})$, we can choose the value of $\delta$ such that the amplitude contributions by $H_s$ and $U$ remain almost unchanged but the phase factor by $H_b$ changes drastically. Consequently, summing over these paths will result in cancelation. The only exception is paths with
\[
H_b(\b{k}_1)=H_b(\b{k}_2)=...=H_b(\b{k}_N)
\]
where the contribution of $H_b$ becomes a trivial global phase factor.
Therefore, we see that the presence of the $H_b$ term simply restricts available paths to those on the constant energy contour of $H_b(\b{k})$.

Now for the surface of a 3D TI without the $H_b$ term, we simply remove the restriction that $\b{k}$ must stay on a constant energy contour. The precessional random walk picture still holds even though there is no semi-classical orbital motion.

\section*{Supplementary Note 3: Difficulties in existing models for the minimum conductivity in 3D TIs}
In this section we address in details why the two currently existing models do not explain the minimum conductivity satisfactorily.

In \cite{tran_5}, the authors showed a resistance peak of $70\ \Omega$ while tuning the gate voltage applied to a $10\ \mathrm{nm}$-thick $\mathrm{Bi}_2\mathrm{Se}_3$ thin film. Considering the $1:8$ aspect ratio of the conduction channel, this resistance converts to a $560\ \Omega$ square resistivity ($\sim50e^2/h$). The authors attributed this conductance to electrons hopping in a bulk impurity band. Charged impurities in $\mathrm{Bi}_2\mathrm{Se}_3$ are believed to have a relatively large Bohr radius $a_B\approx 4\ \mathrm{nm}$, which is comparable with the average spacing between impurities at a typical impurity concentration ($\sim 10^{19}\  \mathrm{cm}^{-3}$). This may result in a considerable hopping amplitude between impurity orbitals and contribute to some conduction if this impurity band is partially occupied. Based on this model, the 2D carrier density contributed by impurities is $n_{2D}<10^{19}\  \mathrm{cm}^{-3}\times10\ \mathrm{nm}=10^{13}\ \mathrm{cm}^{-2}$. To account for the residue conduction, the mobility of such hopping is then greater than $1000\ \mathrm{cm}^2\mathrm{V}^{-1}\mathrm{s}^{-1}$, which is unreasonably high. Based on a similar consideration, Ref. \cite{tran_9} extracted a slightly lower impurity band mobility of $380\ \mathrm{cm}^2\mathrm{V}^{-1}\mathrm{s}^{-1}$ but still seems too high for a hopping mechanism, which should typically be below $1\ \mathrm{cm}^2\mathrm{V}^{-1}\mathrm{s}^{-1}$ \cite{hopping_mobility}. Moreover, hopping electrons should also contribute to Hall coefficient depending on the occupancy of the impurity band. If impurity conduction dominates in the region near the charge neutral point, tuning the surface states shouldn't cause a significant change in Hall coefficient. Although Ref.\cite{tran_5} did not report a Hall coefficient polarity switching, a similar experiment by \cite{tran_4} did report such switching in Ca-doped $\mathrm{Bi}_2\mathrm{Se}_3$ thin films. Ca-doping is expected to induce even higher impurity levels compared to exfoliated single crystals. Therefore impurity band conduction does not seem to be a good explanation of the residual conductivity.

Ref. \cite{tran_6} adopts another explanation which attributes the conductance residue to electron/hole puddles formed when the Fermi level is close to the Dirac point. This model inherits from a similar study in graphene which concludes that the main source of scattering in graphene is unscreened long range Coulomb scattering \cite{trans_gra_thy}. This long range interaction results in a surface potential fluctuation in a relatively large length scale, where electrons can be semi-classically thought to form ``puddles''. However, it has been demonstrated that in the most common 3D TIs such as $\mathrm{Bi}_2\mathrm{Se}_3$, the dominant impurity source is Se vacancies, which is short range and cannot be thought in terms of a semi-classical potential fluctuation. On the other hand, if the minimum conductivity $\sim 5e^2/h$ observed in \cite{tran_6} indeed comes from long range potential fluctuation, a brief estimation reveals that the residue carrier density is $n^\ast\approx10^{12}\ \mathrm{cm}^{-2}$, which corresponds to a potential fluctuation of $120\ \mathrm{meV}$. Those puddle-like residue carriers actually form a lot of mini-pn-junctions and should not be as mobile as uniform carriers. Therefore, the actually required fluctuation is even larger to account for the large $\sigma$. The potential fluctuation on the surface of $\mathrm{Bi}_2\mathrm{Se}_3$ can be directly measured through scanning tunneling spectroscopy, which has already been carried out by several groups. Ref. \cite{pfluctuation_2} did report a typical fluctuation of about $120\ \mathrm{mV}$, but suggested this fluctuation is structural rather than disorder-induced. Moreover, the morphology does not really look like ``puddles'' but rather some ``spikes''. On the other hand, Ref. \cite{pfluctuation_1} reported a much smaller value around $10\ \mathrm{mV}$, suggesting such potential fluctuation is quite sample-dependent and cannot be universally adopted to explain the residue conductivity.